\def \be {\begin{equation}}
\def \ee {\end{equation}}
\def \bea {\begin{eqnarray}}
\def \eea {\end{eqnarray}}
\def \nn {\nonumber}

\def \del {\partial}
\def \dels {\partial\kern-.5em / \kern.5em}
\def \As {{A\kern-.5em / \kern.5em}}
\def \Ds {D\kern-.7em / \kern.5em}

\def \a {\alpha}

\def \dag {\dagger}
\def \g {\gamma}
\def \G {\Gamma}
\def \d {\delta}
\def \eps {\epsilon}

\def \lam {\lambda}
\def \Lam {\Lambda}

\def \th {\theta}
\def \Th {\Theta}

\def \II {I\hspace{-.1em}I\hspace{.1em}}

\def \IIB {\mbox{\II B\hspace{.2em}}}

\def \cL {{\cal L}}

\documentclass[12pt]{article}

\begin{document}

\begin{center}
\hfill hep-th/0010165\\
\vskip .5in

\textbf{\Large \textbf{
Fuzzy Sphere from Matrix Model
}}

\vskip .5in
{\large Pei-Ming Ho}
\vskip 15pt

{\small \em
Department of Physics \\
National Taiwan University \\
Taipei 106 \\
Taiwan}

\vskip .2in
\sffamily{pmho@phys.ntu.edu.tw}
\vspace{60pt}
\end{center}

\begin{abstract}

Using a D0-brane as a probe,
we study the spacetime geometry
in the neighborhood of N D-branes
in matrix theory.
We find that due to fermionic zero modes,
the coordinates of the probe
in the transverse directions
are noncommutative,
and the angular part is a fuzzy sphere. 

\end{abstract}

\setcounter{footnote}{0}

\newpage

\section{Introduction}

The purpose of this paper is to study
the noncommutative nature of spacetime
in the neighborhood of D-branes. 
It was conjectured \cite{JR,HRT} that
the $AdS\times S$ spacetime,
which is the near horizon geometry
of certain brane configurations,
is best described as a noncommutative space,
based on hints from the AdS/CFT duality.
With the help of giant gravitons \cite{MST},
people have also studied the $AdS\times S$ geometry
probed by a graviton,
which can be a string or membrane
depending on the context \cite{HL1,HL2,DJM}.
Although it was first suggested that the spacetime
consists of a q-deformed sphere \cite{JR,HRT},
some evidence was also found for the fuzzy sphere
\cite{BV,HL1,HL2,DJM,JMR1}.
There is also a suggestion of certain equivalence
between two types of noncommutative spheres \cite{JMR2}.

Apart from information from the CFT side,
since D-branes in background NS-NS B field
is known to be noncommutative \cite{NC},
by dualities this implies that
fundamental strings in background R-R B field
become noncommutative \cite{CHK}.
This was indeed shown in the matrix theory \cite{Schiappa}.
Thus we expect that due to the R-R gauge field background,
the $AdS\times S$ spacetime is also noncommutative.

It was pointed out in \cite{BV} that
the AdS/CFT duality can be understood as
a change of basis at the tip of
the Coulomb branch of the DLCQ theory.
By studying the DLCQ theory of the $(2,0)$ field theory,
the $S^4$ part in $AdS_7\times S^4$ was shown
to be a fuzzy sphere in that paper.
In this paper we aim to generalize the derivation
of fuzzy spheres to branes of arbitrary dimensions.
We study the matrix model \cite{BFSS} of a D-particle probing
the spacetime surrounding $N$ coincident D$p$-branes.
(By D-particle we actually mean a graviton
with unit longitudinal momentum in DLCQ M theory.)
We find that the transverse coordinates
of the probe satisfy the algebra of
Snyder \cite{Snyder} and Yang \cite{Yang},
with the angular part defined on a fuzzy sphere
for any even $p$.

We shall focus on the geometry of the space transverse to
$N$ flat D$p$-branes located at the origin.
In the framework of BFSS matrix model \cite{BFSS},
we consider $p=4$ in Sec.\ref{D4},
$p=2,6,8$ in Sec.\ref{D268}.
The case $p=0$ is discussed in Sec.\ref{D0}.
Finally we make some comments in Sec.\ref{remarks}.

\section{D4-branes} \label{D4}

In this section we revisit the case of D4-branes,
for which the fuzzy nature of spacetime
was first discussed in \cite{BV}.

The matrix model for a D-particle in the vicinity of
$N$ D4-branes can be found in \cite{BD}.
\footnote{
The action for $N$ D0-branes and one D4-brane is given in \cite{BD}.
It is straightforward to change it to the action we need
for one D0-brane and $N$ D4-branes.
}
The Lagrangian is composed of three parts
\be \label{D0D4}
\cL=\cL_0+\cL_1+\cL_2,
\ee
where $\cL_0$ is the $U(N)$ SYM theory
on 4+1 dimensions describing $N$ D4-branes,
$\cL_1$ is the SUSY $U(1)$ Lagrangian for the D0-brane,
and the interaction term is
\bea
\cL_2&=&|D_t v^{\rho}|^2+\chi^{\dag}D_t\chi
        -v^{\dag}_{\rho}(X^a-x^a)^2 v^{\rho} \nn\\
     & &-\chi^{\dag}(X^a-x^a)\g_a\chi
        -v^{\dag}_{\rho}(\Th-\psi)^{\rho}\chi
        -|v|^4. \label{L2}
\eea
We have $v^{\rho}$ as a Weyl spinor of $SO(4)$,
and $\chi$ is an $SO(5)$ spinor.
They correspond to the open string stretched between
the D0-brane and $N$ D4-branes,
so they are in the fundamental representation of $U(N)$.
$X$ and $\Th$ are the matrix coordinates for the $N$ D4-branes,
and $x$ and $\psi$ are those for the D0-brane.
The D4-branes are at the origin so we take $X=0$.

For large $x$, one can integrate out $v$ and $\chi$
and read off the metric of D4-brane in
the effective Lagrangian for $x$ \cite{BD}.
For small $x$ we consider the low energy limit in which
we can ignore the kinetic term of $x$.
By varying $\cL$ with respect to $x$, we find
\be \label{x}
x^a=\frac{1}{2|v|^2}\chi^{\dag}\g^a\chi.
\ee
As a change of coordinates,
we can view
\be
r=\frac{1}{2|v|^2}
\ee
as the radial coordinate and
\be
Y^a=\chi^{\dag}\g^a\chi
\ee
as Cartesian coordinates parametrizing $S^4$.
Assuming canonical commutation relations for $\chi$
\be \label{chichi}
\{\chi^{\dag}_{\rho},\chi_{\kappa}\}=\d_{\rho\kappa},
\ee
one finds that the angular coordinates $Y$ satisfy
the algebra of Snyder \cite{Snyder} and Yang \cite{Yang}
\bea
{[Y_a, Y_b ]}&=&iL_{ab}, \label{XX}\\
{[L_{ab}, Y_{c} ]}&=&i\left(\d_{ac}Y_b-\d_{bc}Y_a\right), \label{LX}\\
{[L_{ab}, L_{cd} ]}&=&i\left(\d_{ac}L_{bd}+\d_{bd}L_{ac}
                 -\d_{ad}L_{bc}-\d_{bc}L_{ad}\right) \label{LL},
\eea
where $L_{ab}$ are the generators of $SO(5)$ rotations of the sphere.
This result was first derived in \cite{BV},
and then in \cite{HL1,DJM} in different ways.

Note that the quantization of the system (\ref{chichi}) is
carried out without imposing any additional constraint.
This is in contrast with other derivation of noncommutative space
by imposing BPS constraints \cite{DJM},
or other reduction of the Hilbert space \cite{BV}.

\section{D$p$-branes} \label{D268}

To generalize the discussions above to $p=2,6,8$,
we start with the BFSS matrix model description
of a D0-brane and $N$ D$p$-branes
realized as a bound state of infinitely many D0-branes.
The matrix model has the SYM action
\be \label{action}
S=\int d^{p+1}x \; \mbox{Tr}\left(
\frac{1}{4}[X_{\mu},X_{\nu}]^2
+\frac{1}{2}\bar{\Psi}\G^{\mu}[X_{\mu},\Psi]
\right),
\ee
where $X$'s are the matrix coordinates,
and $\Psi$ is a $9+1$ dimensional Majorana-Weyl spinor
dimensionally reduced to $p+1$ dimensions.
They are
\be \label{ansatz}
X_{\mu}=\left(\begin{array}{cc}
                Z_{\mu}        & y_{\mu} \\
                y^{\dag}_{\mu} & x_{\mu}
          \end{array}\right), \quad
\Psi=\left(\begin{array}{cc}
                \Th        & \th \\
                \th^{\dag} & \psi
           \end{array}\right).
\ee
Here $x$ and $\psi$ are the probe coordinates.
In the temporal gauge $X_0=i\del_0$.
$Z_i=-iD_i$ ($i=1,\cdots,p$) are
covariant derivatives on the dual torus,
and they satisfy the D$p$-brane condition
\be
[Z_{2i-1}, Z_{2i}]=if, \quad i=1,\cdots,p/2,
\ee
where $f$ is a constant depending on $N$,
the longitudinal momentum and the size of the torus.
$Z_a$'s ($a=p+1,\cdots,9$) represent
transverse coordinates of the $N$ D$p$-branes,
so we let $Z_a=0$.
We shall also take $\Th=0$.
In other words, we set the $N$ D-branes
in the ground state as a fixed background.
The open string stretched between the probe
and the $N$ D-branes is represented by $y$ and $\th$,
which are in the fundamental representation
of the gauge group $U(N)$.

Plugging (\ref{ansatz}) into the action (\ref{action}),
we obtain the potential \cite{HLW}
\bea \label{potential}
L&=&\frac{1}{2}|(Z_{\mu}y_{\nu}-Z_{\nu}y_{\mu}
    +y_{\mu}x_{\nu}-y_{\nu}x_{\mu})|^2 \nn\\
 & &+y^{\mu\dag}[Z_{\mu}, Z_{\nu}]y_{\nu}
    -y_{\mu}^{\dag}y_{\nu}y_{\mu}^{\dag}y_{\nu}
    +\frac{1}{2}y_{\mu}^{\dag}y_{\mu}y_{\nu}^{\dag}y_{\nu}
    +\frac{1}{2}y_{\mu}^{\dag}y_{\nu}y_{\nu}^{\dag}y_{\mu} \nn\\
 & &+\bar{\th}\G^{\mu}(Z_{\mu}\th-\th x_{\mu})
    +\bar{\th}\G^{\mu}(y_{\mu}\psi-\Th y_{\mu})
    +(\bar{\psi}y_{\mu}^{\dag}-y_{\mu}^{\dag}\bar{\Th})\G^{\mu}\th.
\eea
For large $x$, the open string modes are heavy,
so one can integrate out the off-diagonal components $y, \th$.
The low energy effective action describes
the probe moving in a curved spacetime.
Given enough supersymmetry,
it is in agreement with supergravity \cite{DKPS}.

However, for small $x$, some open string modes are light,
and they should not be integrated out.
To take into account the light modes,
we first set $x=0$ and find the zero modes,
and then one can adiabatically increase $x$ to a small value
(compared with the string length scale $l_s$).
Given these open string modes,
we shall minimize the potential energy
in order to derive the properties of spacetime
probed by a D0-brane in the low energy limit.

Let us recall some facts about the zero modes \cite{HLW}.
For any $p$ ($p=0, 2,4,6,8$)
there is always a fermionic zero mode.
It is therefore important to keep the fermionic fields.
We will see later that indeed they play a crucial role
in making the spacetime noncommutative.
For $p=2,4,6,8$, denote the fermionic zero mode as
\be \label{zero-f}
\th_{A\a}=\phi_A\chi_{\a}.
\ee
For $p=4$, $\phi_A$ is an $SO(4)$ Weyl spinor on D4-branes.
It is a section of the twisted bundle on the dual torus.
The amplitude of the fermionic zero mode is given by $\chi_m$,
which is an $SO(5)$ spinor.
The bosonic zero mode is
\be \label{zero-b}
y_i=\bar{v}\g_i\phi,
\ee
where $v$ is an $SO(4)$ Weyl spinor.
This can be viewed as a SUSY transformation from
the fermionic zero mode according to
$\d y_{\mu}=\bar{\eps}\G_{\mu}\th$.
Plugging this solution of $\th$ and $y$ into (\ref{potential}),
and integrating ove the dual torus,
one reproduces (\ref{L2}) as the leading order terms.
For $p=2,6,8$, $\phi_A$ is an $SO(p)$ Weyl spinor on D$p$-branes,
and $\chi_m$ is an $SO(9-p)$ spinor.
There is no bosonic zero mode in these cases.

We will ignore the kinetic terms of $x$ and $\psi$ in the action,
assuming that they change very slowly with time.
We will also replace $\th$ by its zero mode solutions at $x=0$.
For $p=2,6,8$, the lowest energy mode of $y$ is given by
\be \label{y}
y_i=v_i^A\phi_A, \quad y_a=0.
\ee
Its mass squared at $x=0$ is of the order $f$ for $p=6,8$.
For $p=2$ the mass squared is negative,
as the D0-brane tends to dissolve into the D2-branes \cite{Taylor}.

{}From (\ref{potential}) we derive the equations of motion for $x$
\be \label{x0}
2|y|^2 x_{\mu}-y_{\mu}^{\dag}y_{\nu}x_{\nu}
-(y_{\nu}^{\dag}x_{\nu})y_{\mu}-\bar{\th}^{\dag}\G_{\mu}\th
-y^{\dag}_{\nu}D_{\nu}y_{\mu}+D_{\nu}y^{\dag}_{\mu}y_{\nu}=0,
\ee
where $\mu,\nu=1,2,\cdots,9$.
Plugging (\ref{zero-f}) and (\ref{y}) into (\ref{x0}),
and integrating over the dual torus,
one finds
\be
2|v|^2 x_a=\bar{\chi}^{\dag}\g_a\chi.
\ee
This again implies (\ref{x}), as in the case $p=4$,
and results in the algebra of an $(8-p)$ dimensional fuzzy sphere.

Strictly speaking, the mass of higher exitations of $y$
are of the same order of magnitude as the lowest energy mode.
Hence it is not a very good approximation to ignore them.
For a generic $y$, we can still integrate (\ref{x0})
over the dual torus for $x_i=0$.
We get
\be \label{x1}
M_{ab}x_b=\bar{\chi}\g_a\chi+z_a,
\ee
where $M_{ab}$ is the integration of
$(2|y|^2\d_{ab}-y_a^{\dag}y_b-y_b^{\dag}y_a)$.
Since $M_{ab}$ is symmetric,
it can be diagonalized by an $SO(p)$ transformation
so that (\ref{x1}) is turned into
\be \label{x2}
M_a x'_a=\bar{\chi}'\g_a\chi'+z'_a.
\ee
Upon quantization, $x'_a$ becomes a coordinate on a fuzzy ellipsoid.
Since all modes of $y$ are heavy for $p=6,8$,
one can also integrate out $y$ (for a given cutoff) and
find effective values for $M_a$ and $z_a$.
Due to the $SO(p)$ symmetry of the system,
we expect that $M_a$ is a constant independent of $a$ and $z_a=0$.
It follows that $x$ is still of the form (\ref{x})
\be \label{xa}
x_a=r\bar{\chi}\g_a\chi,
\ee
and its angular part lives on a fuzzy sphere (\ref{XX})-(\ref{LL}).

\section{D0-branes} \label{D0}

The discussion in the previous section also applies
to the case $p=0$ with constant zero modes.
However, since this case is simpler,
we are able to provide a more explicit derivation.

In the low energy limit we ignore
time derivatives of $x$ and $\psi$ in the action
to derive equations of motion from
the potential (\ref{potential}).
Ignoring also higher order terms of $y$,
we find the equation of motion for $y$
\be \label{y0}
\ddot{y}_a-x_b(y_b x_a-y_a x_b)-\bar{\psi}\G^a\th=0,
\ee
where $a,b=1,\cdots,9$.
This equation of motion has the symmetry
$y_a\rightarrow y_a+\lam x_a$ for constant $x$,
where $\lam$ is any scalar
in the fundamental representation of $U(N)$.
This is in fact a gauge symmetry.
For the background configuration
\be
X_a=\left(\begin{array}{cc}
                0 &  0     \\
                0 & x_a
          \end{array}\right),
\ee
the gauge transformation by
\be
\Lam=\left(\begin{array}{cc}
             0            & \lam \\
             -\lam^{\dag} & 0
           \end{array}\right)
\ee
results in a change of the off-diagonal components
\be
\d X_a=[\Lam, X_a]
      =\left(\begin{array}{cc}
                0              & \lam x_a \\
                \lam^{\dag}x_a & 0
             \end{array}\right).
\ee
We can fix the gauge by imposing
\be \label{yx}
x_a y_a=0.
\ee
Note that $D_a=0$ ($a=1,\cdots,9$) for D0-branes.
Using the equation of motion for $x$ (\ref{x0})
and the gauge fixing condition (\ref{yx}), we find
\be
x_a=\frac{1}{2|y|^2}\bar{\th}\G_a\th.
\ee
This is of the same form as (\ref{x}).
The angular part
\be
Y_a=\bar{\th}\G_a\th
\ee
lives on a fuzzy sphere (\ref{XX})-(\ref{LL}).
Finally we have shown that the fuzzy sphere
appears for all $p=0,2,4,6,8$.

The equation of motion for $\th$ is
\be
\G^a(\th x_a-y_a\psi)=0. \label{th}
\ee
When $x=0$ and $\psi=0$,
equations of motion for $y$ and $\th$ are trivially satisfied.
The zero modes for $y$ and $\th$ are just arbitrary constants.

For a complete analysis,
Gauss' law should be imposed as a constraint of quantization
in the temporal gauge ($A_0=0$).
Time derivatives can not be ignored in the Gauss' law
because the conjugate momenta of $x$
also takes part in the commutation relations.

\section{Remarks} \label{remarks}

The analysis above is valid when $|x|$ is much smaller than $l_s$,
so that the open string zero mode is much lighter than
the oscillations modes and dominates the open string effect.
The string coupling constant $g_s$ should also be very small
to justify our ignoring the closed string effects.
Furthermore, in the low energy limit a D0-brane
can be used to probe structures down to
the scale of the Planck length $g_s^{1/3}l_s$ \cite{DKPS},
thus the fuzzy sphere effect should be important for
physics in the region $g_s^{1/3}l_s<|x|<l_s$.

The derivation above also suggests that
as a perturbative expansion,
it is more efficient to formulate the theory
in terms of the perturbative fields $\tilde{x}$, etc.
defined by
\be
x_a=x_a^{(0)}+\tilde{x}_a,
\ee
where $x^{(0)}$ is given by (\ref{x}).
Since $\tilde{x}$ represents heavier modes of open strings,
one can integrate out $\tilde{x}$ and other heavy modes
to obtain a low energy effective theory for a D0-brane
living on a fuzzy space with coordinates $x_a^{(0)}$.

Our result agrees completely with \cite{BV,HL1} for $S^4$.
(D4-brane here actually means longitudinal M5-brane.)
For the case of $S^2$ (D6-brane), it was proposed \cite{HL1} that
the $AdS_2\times S^2$ space as the near horizon region
of a set of M2 and M5 branes is actually a fuzzy space.
The $S^2$ part is a fuzzy sphere defined by
\be
[Y_a, Y_b]=i\eps_{abc}Y_c.
\ee
This can be viewed as a realization of the algebra
(\ref{XX})-(\ref{LL}) on BPS states satisfying
\be \label{fuzzy}
L_{ab}=\eps_{abc}Y_c.
\ee
These are exactly the states considered in \cite{HL1}
that lead to the suggestion of a fuzzy sphere (\ref{fuzzy}).

The D$p$-branes in type \IIB string theory ($p=$ odd)
are related to the even $p$ cases by T-duality,
thus we also expect fuzzy spaces to appear there.
Maybe one should use the IKKT matrix model \cite{IKKT}
for an analogous discussion.
However, for odd $p$ the angular coordinate $Y$ lives on
an odd dimensional fuzzy sphere.
For even dimensional fuzzy spheres,
the radius squared $R^2=Y^2$ is central
for the representations constructed out of
symmetrized tensor products of Clifford algebras \cite{CLT,HL1}.
Yet for odd dimensions,
we do not know a general construction
of representations for which $R^2$ is central.

In \cite{JR,HRT,HL1,HL2} it was also conjectured that
the $AdS$ part of the spacetime for AdS/CFT dualities
should also be noncommutative.
In our derivation following (\ref{x0}),
we may keep $x_i$ nonzero in the derivation.
The problem is that we do not have
the $SO(9)$ symmetry needed to reach (\ref{x2}).
It is also unclear how the time coordinate
can ever turn out to be noncommutative in this framework.
Perhaps one should start with IKKT matrix model \cite{IKKT}.

\section*{Acknowledgment}

The author thanks Chong-Sun Chu, Hsien-chung Kao and Miao Li
for valuable discussions.
This work is supported in part by
the National Science Council, Taiwan, 
and the Center for Theoretical Physics at National Taiwan University.


\vskip .8cm
\baselineskip 22pt
\begin{thebibliography}{99}
\itemsep 0pt

\bibitem{JR}
A. Jevicki, S. Ramgoolam:
``Noncommutative Gravity From the AdS/CFT Correspondence'',
JHEP 9904: 032 (1999),
hep-th/9902059.

\bibitem{HRT}
P.-M. Ho, S. Ramgoolam, R. Tatar:
``Quantum Spacetime and Finite N Effects in 4D Super Yang-Mill Theories'',
Nucl. Phys. B573: 364-376 (2000),
hep-th/9907145.

\bibitem{MST}
R. C. Myers:
``Dielectric Branes'',
JHEP 9912:022 (1999),
hep-th/9910053.\\
J. McGreevy, L. Susskind, N. Toumbas:
``Invasion of the Giant Gravitons From Anti-De Sitter Space'',
JHEP 0006: 008 (2000),
hep-th/0003075.

\bibitem{HL1}
P.-M. Ho, M. Li:
``Fuzzy Spheres in AdS/CFT Correspondence and
Holography From Noncommutativity'',
to appear in Nucl. Phys. {\bf B},
hep-th/0004072.

\bibitem{HL2}
P.-M. Ho, M. Li:
``Large N Expansion From Fuzzy $AdS_2$'',
to appear in Nucl. Phys. {\bf B},
hep-th/0005268.

\bibitem{DJM}
S. R. Das, A. Jeviski, S. D. Mathur:
``Giant Gravitons, BPS Bounds, and Noncommutativity'',
hep-th/0008088.

\bibitem{BV}
M. Berkooz, H. Verlinde:
``Matrix Theory, AdS/CFT and Higgs-Coulomb Equivalence'',
JHEP 9911:037 (1999),
hep-th/9907100.

\bibitem{JMR1}
A. Jevicki, M. Mihailescu, S. Ramgoolam:
``Noncommutative Spheres and the AdS/CFT Correspondence'',
JHEP 0010: 008 (2000),
hep-th/0006239.

\bibitem{JMR2}
A. Jevicki, M. Mihailescu, S. Ramgoolam:
``Hidden Classical Symmetry in Quantum Spaces
at Roots of Unity: From q-Sphere to Fuzzy Sphere'',
hep-th/0008186.

\bibitem{NC}
A. Connes, M. R. Douglas, A. Schwarz:
``Noncommutative Geometry and Matrix Theory:
Compactification on Tori'',
JHEP 02 (1998) 003,
hep-th/9711162.\\
C.-S. Chu, P.-M. Ho:
``Non-commutative Open String and D-brane'',
Nucl. Phys. {\bf B550} (1999) 151-168,
hep-th/9812219.\\
V. Schomerus:
``D-Branes and Deformation Quantization'',
JHEP 9906: 030 (1999),
hep-th/9903205.\\
N. Seiberg, E. Witten:
``String Theory and Noncommutative Geometry'',
JHEP 9909 (1999) 032,
hep-th/9908142.

\bibitem{CHK}
C.-S. Chu, P.-M. Ho, Y.-C. Kao:
``Worldvoulme Uncertainty Relations for D-Branes'',
Phys. Rev. {\bf D60}: 126003 (1999),
hep-th/9904133.

\bibitem{Schiappa}
R. Schiappa:
``Matrix Strings in Weakly Curved Background Fields'',
hep-th/0005145.

\bibitem{BFSS}
T. Banks, W. Fischler, S. H. Shenker, L. Susskind:
``M Theory as a Matrix Model: a Conjecture'',
Phys. Rev. {\bf D55}: 5112-5128 (1997),
hep-th/9610043.

\bibitem{Snyder}
H. S. Snyder:
``Quantized Space-Time'',
Phys. Rev. 71 (1946) 38.

\bibitem{Yang}
C.-N. Yang:
``On Quantized Space-Time'',
Phys. Rev. 72 (1947) 874.

\bibitem{BD}
M. Berkooz, M. R. Douglas:
``Five Branes in M(atrix) Theory'',
Phys. Lett. {\bf B395} (1997) 196-202,
hep-th/9610236.

\bibitem{HLW}
P.-M. Ho, M. Li, Y.-S. Wu:
``$p-p'$ Strings in M(atrix) Theory'',
Nucl. Phys. {\bf B525} (1998) 146-162.
hep-th/9706073.

\bibitem{DKPS}
M. R. Douglas, D. Kabat, P. Pouliot, S. H. Shenker:
``D-branes and Short Distances in String Theory'',
Nucl. Phys. {\bf B485}: 85-127 (1997),
hep-th/9608024.

\bibitem{Taylor}
W. Taylor IV:
``Adhering 0-branes to 6-branes and 8-branes'',
Nucl. Phys. {\bf B508} (1997) 122,
hep-th/9705116.

\bibitem{IKKT}
N. Ishibashi, H. Kawai, Y. Kitazawa, A. Tsuchiya:
``A Large N Reduced Model as Superstring'',
Nucl. Phys. {\bf B498}: 467-491 (1997),
hep-th/9612115.

\bibitem{CLT}
J. Castelino, S. Lee, W. Taylor IV:
``Longitudinal Five Branes as Four Spheres in Matrix Theory'',
Nucl. Phys. {\bf B526}: 334-350 (1998),
hep-th/9712105.

\end{thebibliography}
\end{document}